\documentclass[onecolumn,floatfix,aps,nofootinbib, showpacs, showkeys, 10pt]{revtex4-2}

\usepackage[dvips]{epsfig}
\usepackage[english]{babel}
\usepackage{bbm}
\usepackage{verbatim}
\usepackage{array}
\usepackage{amsmath}
\usepackage{multirow}
\usepackage{amsfonts}
\usepackage{amssymb}
\usepackage{hyperref}
\usepackage{leading}
\usepackage[dvipsnames]{xcolor}
\hypersetup{colorlinks=true,linkbordercolor=Blue,linkcolor=Blue, citecolor=Blue}
\usepackage{subeqnarray}
\usepackage{setspace}
\usepackage{placeins}
\usepackage{float}
\usepackage{indentfirst} 

\usepackage{gensymb}

\usepackage{bbold}
\def\I{\openone}
\def\openone{\mathbb I}
\def\0{\mathbb{0}}

\def\p{\mbox{\boldmath$\displaystyle\boldsymbol{p}$}}
\newcommand{\gdualn}[1]{\overset{\:{}^{{}^{\boldsymbol{\neg}}}}{\smash[t]{#1}}}
\def\R{\mbox{\boldmath$\displaystyle\mathbb{R}$}}

\begin{document}

\title{Exploiting new classes of mass-dimension one fermions}

\author{Rodolfo Jos\'{e} Bueno Rogerio$^{1}$ and Luca Fabbri$^{2}$}
\affiliation{$^{1}$Institute of Physics and Chemistry, Federal 
University of Itajub\'{a}, Itajub\'{a}, Minas Gerais, 37500-903, BRAZIL \\ 
$^{2}$DIME Sez. Metodi e Modelli Matematici, Universit\`{a} di
Genova, via all'Opera Pia 15, 16145 Genova, ITALY}


\begin{abstract}
{\bf Abstract} In this paper, we employ the machinery firstly shown in \cite{dharamnewfermions,dharamboson}, to obtain a new class of spin-half mass-dimension one fermions. Such spinors, after an appropriated dual structure examination, may serve as expansion coefficients of a local field.
\end{abstract}


\maketitle

\section{Introduction}\label{intro}
According to Lounesto's classification\cite{lounestolivro}, spin-$\frac{1}{2}$ spinor fields can be split into two major classes, regular and singular, respectively containing Dirac spinor fields and flag-dipole spinor fields \cite{dualtipo4,chengflagdipole}; the last class in particular can be subject to further splitting, giving rise to flagpoles (among which Majorana spinors) and dipoles (among which Weyl spinors). Dirac, Majorana and Weyl spinors are the pillars upon which we build quantum field theory and the standard model of particle physics, and they are all very well-known. However, from what we can understand from the method followed by Lounesto, the whole spinor classification is capable of assessing the physical properties of many more spinorial structures than those that are normally considered to be fundamental (and that is Dirac, Majorana and Weyl spinors). In order to avoid leaving fundamental information behind, it may therefore be wise to extend investigations so as to include all possible spinor structures.

Quite recently, in Refs \cite{dharamnewfermions, dharamboson}, a new mathematical tool has been investigated, the so-called square root of $4\times4$ identity matrix, which provides a mechanism to define the expansion coefficients of spin one-half fermionic fields \cite{dharamnewfermions}. The given method allows to construct new classes of fermionic fields, based on eigenvectors of a linearly-independent set of re-arrangements of the Clifford matrices, which are however characterized for their being of mass-dimension one. They are therefore a sort \emph{bosonic fermions}. They are essentially single-helicity spinors that do not necessarily satisfy the Dirac equations, called Elko \cite{mdobook}. By construction, they are naturally neutral, and thus potential candidates for dark matter \cite{mdobook}. An intrinsic feature of mass-dimension one fermions is that their spin-sums are not Lorentz-covariant \cite{jcap, chengflagdipole, dharamnewfermions}. Consequently, the associated propagators are non-local, but endowed with a \emph{preferred} direction. By introducing the $\tau$-deformation for the adjoint structure dual \cite{aaca, vicinity}, one can easily check that Lorentz-invariance can be restores \cite{chengflagdipole, rodolfosingularconnection}. The Lorentz-invariant formulation is obtained by removing the breaking term, namely $\mathcal{G}$ in the spin-sum. From the physical point of view, the removal of $\mathcal{G}$ via $\tau$-deformation is mathematically well-posed.

The algorithm describe above, about the square root of unity, may then provide new candidates to populate the mass-dimension one sector, but it can also give rise to yet another new entity, namely spin-half bosons \cite{dharamboson}. They are therefore \emph{fermionic bosons}.

In the present paper, however, we would leave fermionic bosons aside, focusing on the new type of bosonic fermions. However, we are not going to focus on Elko, but rather on a type of spinors that cannot be interpreted due to the fact that they do not hold real eigenvalues under the action of $\mathcal{C}$ operator. In fact, one could allow some couplings that are not allowed for Elko fields to to their being neutral. Hence, the spinors introduced here are new entities belonging to the class of flag-pole spinors within the Lounesto classification \cite{lounestolivro}.

The paper is organized as follows: In Sect. \ref{sect1} and \ref{sectdual} we revisit the linearly-independent square root of the identity matrix and introduce the eigenvectors of the $\Gamma_{12}$ set, exploiting the appropriated adjoint structure. In Sect.\ref{sectpropagator}, we define the quantum field operators, built the propagator and compute the Hamiltonian.
\section{The background and the new spinors}\label{sect1}
The mechanism is based on the linearly-independent square roots of identity \cite{schweber}, given by
\begin{eqnarray}
&&\I  \nonumber\\
&&i\gamma_1 \;\;\; i\gamma_2 \;\;\; i\gamma_3 \;\;\; \gamma_0 \nonumber\\
&&i\gamma_2\gamma_3 \;\;\; i\gamma_3\gamma_1 \;\;\; i\gamma_1\gamma_2 \;\;\; \gamma_0\gamma_1 \;\;\; \gamma_0\gamma_2\;\;\; \gamma_0\gamma_3 \\
&&i\gamma_0\gamma_2\gamma_3 \;\;\; i\gamma_0\gamma_1\gamma_3 \;\;\; i\gamma_0\gamma_1\gamma_2 \;\;\; \gamma_1\gamma_2\gamma_3  \nonumber\\
&&i\gamma_0\gamma_1\gamma_2\gamma_3 \nonumber
\end{eqnarray}
denoting the above set of gamma matrices by $\Gamma_{\ell}$, $\ell=1,\cdots, 16$, in which $\Gamma_1=\I$ and $\Gamma_{16}=i\gamma_0\gamma_1\gamma_2\gamma_3$. The extra factor $i$ ensures that $\Gamma^{2}_{\ell}=+\I$ providing real eigenvalues. 
The above representation is irreducible \cite{schweber}. 

We start by exploiting the eigenvectors of $\Gamma_{12} = i\gamma_0\gamma_2\gamma_3$ set, from \cite{dharamboson}, given (in the rest-frame referential) by
\begin{eqnarray}\label{spinors1}
\lambda_1(\textbf{0}) = \sqrt{m}\left(\begin{array}{c}
1 \\ 
0 \\ 
0 \\ 
1
\end{array} \right),\quad \lambda_2(\textbf{0}) = \sqrt{m}\left(\begin{array}{c}
-1 \\ 
0 \\ 
0 \\ 
1
\end{array} \right),
\end{eqnarray}
and 
\begin{eqnarray}\label{spinors2}
\lambda_3(\textbf{0}) = \sqrt{m}\left(\begin{array}{c}
0 \\ 
1 \\ 
1 \\ 
0
\end{array} \right), \quad \lambda_4(\textbf{0}) = \sqrt{m}\left(\begin{array}{c}
0 \\ 
-1 \\ 
1 \\ 
0
\end{array} \right).
\end{eqnarray}
Where we have defined these spinors as rest spinors $\lambda_i(\textbf{0})$. To obtain the spinors for an arbitrary momentum, $\lambda_i(\textbf{p})= \kappa\lambda_i(\textbf{0})$, we act with the $(1/2,0)\oplus(0,1/2)$ boost operator
\begin{eqnarray}
\kappa = \sqrt{\frac{E+m}{2m}}\left(\begin{array}{cc}
\I+ \frac{\vec{\sigma}\cdot\vec{\textbf{p}}}{E+m} & 0 \\ 
0 & \I- \frac{\vec{\sigma}\cdot\vec{\textbf{p}}}{E+m}
\end{array} \right),
\end{eqnarray} 
in which $\boldsymbol{\sigma}$ stands for the Pauli matrices.
 
Since we have defined the momentum-dependent spinors, we may next verify the $\lambda$ spinor dynamics and its behaviour under action of the $\mathcal{C}$, $\mathcal{P}$ and $\mathcal{T}$ discrete symmetries. 

We start inspecting the action of the Dirac operator, yielding the following relations
\begin{eqnarray}
&&\gamma_{\mu}p^{\mu}\lambda_1(\textbf{p}) = m \lambda_3(\textbf{p}),  \quad  \gamma_{\mu}p^{\mu}\lambda_3(\textbf{p}) = m \lambda_1(\textbf{p}),
\\
&&\gamma_{\mu}p^{\mu}\lambda_2(\textbf{p}) = -m \lambda_4(\textbf{p}), \quad \gamma_{\mu}p^{\mu}\lambda_4(\textbf{p}) = -m \lambda_2(\textbf{p}).
\end{eqnarray}
As we can see, the introduced spinors do not satisfy a first-order equation in momentum space. However, the action of $\gamma_{\nu}p^{\nu}$ in the above relations induce the following relation
\begin{eqnarray}
(p_{\mu}p^{\mu}-m^2)\lambda_i(\p)=0,
\end{eqnarray}
thus, as one can see $\lambda_i(\p)$ spinors fulfill the Klein-Gordon equation. The above results can be displayed in the following fashion
\begin{eqnarray}
&&\mathcal{P}\lambda_1(\textbf{p}) =  \lambda_3(\textbf{p}),  \quad  \mathcal{P}\lambda_3(\textbf{p}) =  \lambda_1(\textbf{p}), \label{p1}
\\
&&\mathcal{P}\lambda_2(\textbf{p}) = - \lambda_4(\textbf{p}), \quad \mathcal{P}\lambda_4(\textbf{p}) = - \lambda_2(\textbf{p}).\label{p2}
\end{eqnarray}
in which, $ \mathcal{P} = m^{-1}\gamma_{\mu}p^{\mu}$, as noticed in \cite{speranca}. 

With respect to the charge-conjugation $(\mathcal{C}=\gamma_2\mathcal{K})$ and time-reversal operator ($\mathcal{T} =  i\gamma_5\mathcal{C}$) we have
\begin{eqnarray}
&&\mathcal{C}\lambda_1(\p) = -i\lambda_1(\p) \quad \mathcal{C}\lambda_2(\p) =  i\lambda_2(\p),
\\
&&\mathcal{C}\lambda_3(\p) =  i\lambda_3(\p) \quad \mathcal{C}\lambda_4(\p) = -i \lambda_4(\p),
\end{eqnarray}
and 
\begin{eqnarray}
&&\mathcal{T}\lambda_1(\p) = -\lambda_2(\p) \quad \mathcal{T}\lambda_2(\p) =  \lambda_1(\p),
\\
&&\mathcal{T}\lambda_3(\p) =  \lambda_4(\p) \quad \mathcal{T}\lambda_4(\p) = -\lambda_3(\p).
\end{eqnarray}
With the previous results at hand, we are able to compute $\mathcal{P}^2 = +\I$, $\mathcal{C}^2=+\I$, $\mathcal{T}^2=-\I$, $(\mathcal{C}\mathcal{P}\mathcal{T})^2=+\I$, holding similarity with other mass-dimension one fermions  \cite{dharamnewfermions}.

\section{Defining the dual structure}\label{sectdual}
At this point we move into a more detailed analysis concerning the dual structure definition \cite{juliodual,nondirac}.

Under the Dirac dual structure, the spinors presented in \eqref{spinors1} and \eqref{spinors2}, hold a null norm
\begin{eqnarray}
\bar{\lambda}_j(\p)\lambda_j(\p) = 0, \quad\mbox{for all $j$}.
\end{eqnarray}
So, following the same algorithm in the recent literature \cite{aaca}, we define the dual structure in its most general form
\begin{eqnarray}
\stackrel{\neg}{\lambda}_{j}(\textbf{p}) = [\mathcal{O}\lambda_{j}(\textbf{p})]^{\dag}\gamma_0.
\end{eqnarray}
in which the $\mathcal{O}$ operator has the following structure
\begin{eqnarray}
\mathcal{O} = \frac{1}{2m}[\lambda_1(\p)\bar{\lambda}_1(\p)-\lambda_2(\p)\bar{\lambda}_2(\p)+\lambda_3(\p)\bar{\lambda}_3(\p)-\lambda_4(\p)\bar{\lambda}_4(\p)].
\end{eqnarray}
It is readily seen that $\mathcal{O}^2=\I$ and $\mathcal{O}^{-1}=\mathcal{O}$. The action of the $\mathcal{O}$ operator over the introduced spinors reads
\begin{eqnarray}
\mathcal{O}\lambda_1(\p) = \lambda_3(\p), \quad \mathcal{O}\lambda_2(\p) = \lambda_4(\p)
\\
\mathcal{O}\lambda_3(\p) = \lambda_1(\p), \quad \mathcal{O}\lambda_4(\p) = \lambda_2(\p).
\end{eqnarray} 
A straightforward calculation, after the dual structure is settled, provides
\begin{eqnarray}
&&\stackrel{\neg}{\lambda}_{1}(\textbf{p}) = \lambda^{\dag}_3(\textbf{p})\gamma_0, \quad \stackrel{\neg}{\lambda}_{2}(\textbf{p}) = \lambda^{\dag}_4(\textbf{p})\gamma_0,\label{dualamb1}
\\
&&\stackrel{\neg}{\lambda}_{3}(\textbf{p}) = \lambda^{\dag}_1(\textbf{p})\gamma_0, \quad \stackrel{\neg}{\lambda}_{4}(\textbf{p}) = \lambda^{\dag}_2(\textbf{p})\gamma_0, \label{duallamb4}
\end{eqnarray}
with orthonormal relations yielding a real and invariant norm
\begin{eqnarray}
\label{orto1}\stackrel{\neg}{\lambda}_{j}(\textbf{p})\lambda_{j}(\textbf{p}) = +2m \quad\mbox{for $j=1,3$}
\\
\label{orto2}\stackrel{\neg}{\lambda}_{j}(\textbf{p})\lambda_{j}(\textbf{p}) = -2m \quad\mbox{for $j=2,4$}
\end{eqnarray}
With the appropriated dual structure at hands, one may compute the spin-sum, obtaining
\begin{eqnarray}\label{spinsum13}
\sum_{j=1,3} \lambda_j\stackrel{\neg}{\lambda}_{j}(\textbf{p}) = m(\I + \mathcal{M}(p, E)),
\end{eqnarray}
and 
\begin{eqnarray}\label{spinsum24}
\sum_{j=2,4} \lambda_j\stackrel{\neg}{\lambda}_{j}(\textbf{p}) = -m(\I - \mathcal{M}(p, E)),
\end{eqnarray}
in which $\mathcal{M}(p, E)$ reads
\begin{eqnarray}
\mathcal{M}(p, E) =
 \left( \begin{array}{cccc}
0 & 0 & \frac{(E+m+p_z)p_x}{m(E+m)} & 1+\frac{p_x(p_x-ip_y)}{m(E+m)} \\ 
0 & 0 & 1+\frac{p_x(p_x+ip_y)}{m(E+m)} & \frac{(E+m-p_z)p_x}{m(E+m)} \\ 
\frac{-(E+m-p_z)p_x}{m(E+m)} & 1+\frac{p_x(p_x-ip_y)}{m(E+m)} & 0 & 0 \\ 
1+\frac{p_x(p_x+ip_y)}{m(E+m)} & \frac{-(E+m+p_z)p_x}{m(E+m)} & 0 & 0
\end{array}  \right).
\end{eqnarray}
The following completeness relation
\begin{equation}
\frac{1}{2 m}
\left[\sum_{j=1,3} {\lambda}_j(\p) \gdualn{\lambda}_j(\p) - 
\sum_{j=2,4} {\lambda}_j(\p) \gdualn{\lambda}_j(\p) \right]= \I
\end{equation}
is also provided. Note that $\mathcal{M}(p, E)$ explicitly breaks the Lorentz covariant structure. Thus, we are forced to re-examine the dual structure in such a way that the spin-sums as computed above become Lorentz invariant. It turns out that it happens only if one multiplies the Lorentz-violating piece in the spin-sums by a parameter involving the inverse of the spin-sums themselves. The new adjoint structure (henceforth indicated as $\tilde{\lambda}$) must be written in terms of the inverse of the matrix $(\I \pm \mathcal{M}(p, E))$. Having established this much, we define the new adjoint as
\begin{eqnarray}
\tilde{\lambda}_j(\textbf{p}) =  \stackrel{\neg}{\lambda}_{j}(\textbf{p})(\I + \mathcal{M}(p, E))^{-1} \quad \mbox{for $j=1,3$}, \label{newdual1}
\\
\tilde{\lambda}_j(\textbf{p}) =  \stackrel{\neg}{\lambda}_{j}(\textbf{p})(\I - \mathcal{M}(p, E))^{-1} \quad \mbox{for $j=2,4$}, \label{newdual2}
\end{eqnarray}   
yielding the following spin-sums
\begin{eqnarray}
\sum_{j=1,3} \lambda_j\stackrel{\neg}{\lambda}_{j}(\textbf{p}) = m(\I + \mathcal{M}(p, E))(\I + \mathcal{M}(p, E))^{-1},
\end{eqnarray}
and 
\begin{eqnarray}
\sum_{j=2,4} \lambda_j\stackrel{\neg}{\lambda}_{j}(\textbf{p}) = -m(\I - \mathcal{M}(p, E))(\I - \mathcal{M}(p, E))^{-1},
\end{eqnarray} 
and they are Lorentz-invariant. A quick inspection reveals that the right-hand side of equations \eqref{spinsum13} and \eqref{spinsum24} do not admit inverse, that is $\det (\I\pm\mathcal{M}(p, E)) =0$.
Bearing in mind that $\mathcal{M}^2(p, E) = \I$ and that the eigenvalues of $\I$ and $\mathcal{M}(p, E)$ are equal to $\pm 1$, we are able to compute the inverse by applying the $\tau$-deformation algorithm presented in \cite{vicinity}, furnishing
\begin{eqnarray}
&&(\I + \mathcal{M}(p, E))^{-1} =\frac{\I - \tau\mathcal{M}(p, E)}{1-\tau^2}, 
\\
&&( \I - \mathcal{M}(p, E))^{-1} = \frac{\I + \tau\mathcal{M}(p, E)}{1-\tau^2}.
\end{eqnarray}
Combining equations \eqref{dualamb1}-\eqref{duallamb4} with \eqref{newdual1} and \eqref{newdual2}, we are able to write the dual structure in the following fashion
\begin{eqnarray}
&&\tilde{\lambda}_{1}(\textbf{p}) =2\stackrel{\neg}{\lambda}_1(\textbf{p})\bigg(\frac{\I - \tau\mathcal{M}(p, E)}{1-\tau^2}\bigg),
\\
&&\tilde{\lambda}_{2}(\textbf{p}) = 2\stackrel{\neg}{\lambda}_2(\textbf{p})\bigg(\frac{\I + \tau\mathcal{M}(p, E)}{1-\tau^2}\bigg),
\\
&&\tilde{\lambda}_{3}(\textbf{p}) = 2\stackrel{\neg}{\lambda}_{3}(\textbf{p}) \bigg(\frac{\I - \tau\mathcal{M}(p, E)}{1-\tau^2}\bigg),
\\
&&\tilde{\lambda}_{4}(\textbf{p}) = 2\stackrel{\neg}{\lambda}_{4}(\textbf{p}) \bigg(\frac{\I + \tau\mathcal{M}(p, E)}{1-\tau^2}\bigg),
\end{eqnarray}
in which the constant multiplicative factor $2$ is necessary to keep the relations \eqref{orto1} and \eqref{orto2} unchanged. After such a redefinition, the spin-sums finally read
\begin{eqnarray}
\sum_{j=1,3} \lambda_j(\textbf{p})\tilde{\lambda}_{j}(\textbf{p}) = 2m\I,
\end{eqnarray}
and 
\begin{eqnarray}
\sum_{j=2,4} \lambda_j(\textbf{p})\tilde{\lambda}_{j}(\textbf{p}) = -2m\I.
\end{eqnarray}
Clearly they are Lorentz-invariant quantities. As can be seen in the current literature, the mechanism described above is the price to pay to establish the invariance of spin-sums, through the redefinition of the dual structure.
A very important point regarding the construction performed in this work is the direct observation of the structure of the eigenvectors of $\Gamma_{\ell}$ and its classification within Lounesto classification. As far as we know, see \cite[and references therein]{rjfermionicfield, rodolfoconstraints}, only a specific subclass of Lounesto's class 2 is composed of spinors that lead to a theory of local quantum fields. Thus, if any of the eigenvectors of $\Gamma_{\ell}$ belong to this subclass, as well as \cite{dharamboson}, we are automatically working in the scope of a local theory, otherwise, as shown in this work and also in \cite{mdobook, dharamnewfermions}, a careful analysis of the dual structure must be done in order to ensure locality. Thus, the roadmap for the presented construction (and also for future constructions) should be based on the right appreciation of the dual structure --- leading to a Lorentz invariant and non-vanishing norm  and also a Lorentz invariant spin sums.

\section{The propagator and Fermi Statistics}\label{sectpropagator}
Having settled the algebraic structure, we now move to investigate the quantum structure of the field in its general form. With $\lambda_i(\p)$ spinors as expansion coefficients we have
\begin{equation}
\mathfrak{b}(x) \stackrel{\mathrm{def}}{=}
\int\frac{\mbox{d}^3 p}{(2\pi)^3}
\frac{1}{\sqrt{2 \upsilon E(\p)}}
\bigg[
\sum_{j=1,3} {a}_j(\p)\lambda_j(\p) e^{-i p\cdot x}+ \sum_{j=2,4} b^\dagger_j(\p)\lambda_j(\p) e^{i p\cdot x}\bigg],\label{eq:fieldb}
\end{equation}
and 
\begin{equation}
\gdualn{\mathfrak{b}}(x) \stackrel{\mathrm{def}}{=}
\int\frac{\mbox{d}^3 p}{(2\pi)^3}
\frac{1}{\sqrt{2 \upsilon E(\p)}}
\bigg[
\sum_{j=1,3} a^\dagger_j(\p)\tilde{\lambda}_j(\p) e^{i p\cdot x}+ \sum_{j=2,4} b_j(\p)\tilde{\lambda}_j(\p) e^{-i p\cdot x}\bigg],
\end{equation}
as its adjoint. The $\upsilon$ parameter is left free to be fixed later times, at our convenience. Such a factor is related to the type of the field we are considering: if we are dealing with the field which obeys only a second-order equation (as mass-dimension one fermions do), then $\upsilon = m$; otherwise, if one is handling a field which fulfills the first-order equation (as for the Dirac spinors) one sets $\upsilon=1$. For now, we are not going to impose any relationship between the particle creation/annihilation operators. We are going to look for such relationships during the construction of the amplitude of propagation. 

As is well known, the fermionic statistic are written as 
\begin{equation}
\left\{a_i(\p),a^\dagger_j(\p)\right\} =(2 \pi)^3 \delta^3(\p-\p^\prime)\delta_{ij}, \quad \left\{a_i(\p), a_j(\p^\prime)\right\} = 0 =
 \left\{a^\dagger_i(\p), a^\dagger_j(\p^\prime)\right\}, \label{anticomutador}
\end{equation}
whereas the bosonic statistics are 
\begin{equation}
\left[a_i(\p),a^\dagger_j(\p)\right] =(2 \pi)^3 \delta^3(\p-\p^\prime)\delta_{ij}, \quad \left[a_i(\p), a_j(\p^\prime)\right] = 0 =
 \left[a^\dagger_i(\p), a^\dagger_j(\p^\prime)\right]. \label{comutador}
\end{equation}
Similar relations are expected for $b_i(\p)$ and $b_i^\dagger(\p)$ operators. 

To determine the statistics for the $\mathfrak{b}(x)$ and $\gdualn{b}(x)$ we follow a similar program as scrutinized in \cite{dharamfermijmpd}, where the authors looked for the main aspects of causality and Fermi statistics for mass-dimension one fermions. We consider two events, $x$ and $x'$, and note that the amplitude to propagate from $x$ to $x'$ is usually written in the following form
\begin{align}
i\mathcal{D}(x- x^\prime)   =  \xi \Big(\underbrace{\langle\hspace{3pt}\vert
\mathfrak{b}(x^\prime)\gdualn{\mathfrak{b}}(x)\vert\hspace{3pt}\rangle \theta(t^\prime-t)
\pm  \langle\hspace{3pt}\vert
\gdualn{\mathfrak{b}}(x) \mathfrak{b}(x^\prime)\vert\hspace{3pt}\rangle \theta(t-t^\prime)}_{\langle\hspace{4pt}\vert \mathfrak{T} ( \mathfrak{b}(x^\prime) \gdualn{\mathfrak{b}}(x)\vert\hspace{4pt}\rangle}\Big)\label{amplitude1}.
\end{align}
One must keep in mind the plus sign holding for bosons and the minus sign for fermions \cite{greiner1990relativistic} and that $\mathfrak{T}$ is the time ordering operator. At this stage, it is worth to re-write the two Heaviside step functions of equation (\ref{amplitude1}) in their integral form
\begin{align}
\theta(t^\prime-t) &=  -\frac{1}{2\pi i}\int\text{d}\omega
\frac{e^{i \omega (t^\prime-t)}}{\omega- i \epsilon}, \\
\theta(t-t^\prime) &=  -\frac{1}{2\pi i}\int\text{d}\omega
\frac{e^{i \omega (t-t^\prime)}}{\omega- i \epsilon},
\end{align}
in which $\epsilon,\omega\in\R$. Now, we are able to write the amplitude of propagation 
\begin{eqnarray}\label{propagador666}
i\mathcal{D}(x- x^\prime)
= -\xi \mathop{\mathrm{lim}}\limits_{\epsilon\rightarrow 0^+}
 \int\frac{\mathrm{d}^3p}{(2\pi)^3}\frac{1}{2mE(\boldsymbol{p})}
\int\frac{\mathrm{d}\omega}{2\pi i}&\times &
\bigg[\frac{\sum_{j=1,3}\lambda(\p)\tilde\lambda(\p)}{\omega - i\epsilon}\mathrm{e}^{i(\omega-E(\boldsymbol{p})(t^\prime -t)}\,\mathrm{e}^{i 
\boldsymbol{p}.(\mathbf{x}^\prime-\mathbf{x})}\nonumber\\
&\pm&
\frac{\sum_{j=2,4}\lambda(\p)\tilde\lambda(\p)}{\omega - i\epsilon}\mathrm{e}^{-i(\omega-E(\boldsymbol{p})(t^\prime -t)}\,\mathrm{e}^{i
\boldsymbol{p}.(\mathbf{x}^\prime-\mathbf{x})}\bigg].
\end{eqnarray} 

By employing these results, substituting $\omega \to p_0 = -\omega+E(\p)$ in the first term and  $\omega \to p_0 = \omega- E(\p)$ in the second term and using the spin-sums defined above, equation \eqref{propagador666} can be written as
\begin{eqnarray}\label{general-propagator-form}
i\mathcal{D}(x- x^\prime)
= i\xi \mathop{\mathrm{lim}}
\limits_{\epsilon\rightarrow 0^+}
\int \frac{\mathrm{d}^4 p}{(2\pi)^4} \frac{1}{2 mE(\boldsymbol{p})} 
\mathrm{e}^{-i p_\mu(x^{\prime\mu} - x^\mu)}
\bigg[ \frac{ \sum_{j=1,3}\lambda(\p)\tilde\lambda(\p)}{E(\p) + p_0 - i\epsilon}
\pm
\frac{\sum_{j=2,4}\lambda(-\p)\tilde\lambda(-\p)} {E(\p) - p_0 - i\epsilon}\bigg].\nonumber\\
\end{eqnarray}
hence furnishing
\begin{eqnarray}
&&\mathcal{D}(x-x^{\prime}) = 
\nonumber\\
&&\xi\bigg(\int \frac{d^4p}{(2\pi)^4}\frac{1}{2mE(\p)}e^{-ip_{\mu}x^{\mu}}\sum_{j=1,3} \lambda_j(\textbf{p})\tilde{\lambda}_{j}(\textbf{p}) + \beta \int \frac{d^4p}{(2\pi)^4}\frac{1}{2mE(\p)}e^{-ip_{\mu}x^{\mu}}\sum_{j=2,4} \lambda_j(-\textbf{p})\tilde{\lambda}_{j}(-\textbf{p})\bigg).
\end{eqnarray} 
The introduced parameter $\beta$ stands for a real constant being $\beta = +1$ for bosons and $\beta = -1$ for fermions. After a straightforward calculation, we have
\begin{eqnarray}\label{45}
\mathcal{D}(x-x^{\prime}) = \xi\int \frac{d^4p}{(2\pi)^4}\frac{1}{2mE(\p)}e^{-ip_{\mu}x^{\mu}}\bigg[\frac{2m\I}{E(\p)-\sqrt{\p^2+m^2}-i\epsilon}+ \frac{(-2m)\I\beta}{E(\p)+\sqrt{\p^2+m^2}-i\epsilon}\bigg],
\end{eqnarray} 
the only relevant physical result comes from the choice $\beta = -1$. This last observation, is equivalent to the choice \eqref{anticomutador}. After some mathematical manipulations in equation \eqref{45}, we obtain
\begin{eqnarray}
\mathcal{D}(x-x^{\prime}) = 2\xi\int \frac{d^4p}{(2\pi)^4}e^{-ip_{\mu}x^{\mu}} \frac{\I}{p_{\mu}p^{\mu}-m^2+i\epsilon}.\label{eq:AmplitudeWithXi}
\end{eqnarray}
Now, to fix the factor $\xi$, we normalize the integral of $\mathcal{D}(x-x^{\prime})$ over all possible $(x-x^{\prime})$ space-time (which corresponds to the amplitude for the particle to be found anywhere in the universe), so that
\begin{eqnarray}
 2\xi \int \frac{d^4p}{(2\pi)^4}(2\pi)^4\delta(p^{\mu})e^{-ip_{\mu}x^{\mu}} \frac{\I}{p_{\mu}p^{\mu}-m^2+i\epsilon}=1.
\end{eqnarray}
Then the above equation translates into
\begin{eqnarray}
2\xi\frac{1}{-m^2+i\epsilon}=1,
\end{eqnarray}
and thus the normalisation of the $\xi$ factor (taking the limit $\epsilon\rightarrow 0$) is
\begin{equation}
\xi = -\frac{m^2}{2}.
\end{equation}
This gives
\begin{equation}
\mathcal{D}(x- x^\prime)  =  -m^2 \int\frac{\text{d}^4 p}{(2 \pi)^4}\,
e^{-i p_\mu(x^{\prime\mu}-x^\mu)}
\frac{\I}{p_\mu p^\mu -m^2 + i\epsilon},
\end{equation}
so that
\begin{equation}
\left(\partial_{\mu^\prime} \partial^{\mu^\prime} \I + m^2\I\right)
S_{\textrm{FD}}(x^\prime-x)  =   -\delta^4(x^\prime - x).
\end{equation}
The Feynman-Dyson propagator is finally
\begin{align}\label{propagatormdofinal}
S_{\textrm{FD}}(x^\prime-x) & \stackrel{\textrm{def}}{=}  -\frac{1}{m^2} 
\mathcal{D}(x- x^\prime)\nonumber\\
 &= \int\frac{\text{d}^4 p}{(2 \pi)^4}\,
e^{-i p_\mu(x^{\prime\mu}-x^\mu)}
\frac{\I}{p_\mu p^\mu -m^2 + i\epsilon}.
\end{align} 
identical to that of a scalar Klein-Gordon field. Thus, such results allow us to write the Lagrangian density
\begin{equation}
\mathfrak{L}(x) =\partial^\mu\gdualn{\mathfrak{b}}\,\partial_\mu {\mathfrak{b}}(x) - m^2 \gdualn{\mathfrak{b}}(x) \mathfrak{b}(x).\label{lagrangiana}
\end{equation} 
Following the arguments presented in pages 500 and 502 from \cite{weinberg1}, Dirac, Weyl or Majorana spinors obey a first-order derivative field equation. Such a feature implies a propagator that for large momentum is proportional to $p^{-1}$. This asymptotic behaviour of the associated propagator results in the fact that mass-dimension must be $3/2$. However, note the unique kinematic operator that is satisfied by $\lambda$ spinor is the Klein-Gordon equation, which is a second-order derivative equation. For this specific case, for a large momentum, the propagator is proportional to $p^{-2}$, contrasting the previous cases. Thus, we are led to conclude that the mass-dimension of the $\mathfrak{b}$ field is 1, rather than 3/2. Such a result is in agreement with other mass-dimension one fermions, as reported in \cite{mdobook, dharamnewfermions}.

We have derived, rather than assumed, the Fermi statistics for the introduced fields.
Once the field equation was verified, one can easily determinate the momentum conjugate to $\mathfrak{b}(x)$
\begin{equation}
\boldsymbol{\pi}(x) = \frac{\partial \mathfrak{L}(x)}
{\partial {\dot{\mathfrak{b}}(x)}} = \frac{\partial \gdualn{\mathfrak{b}}(x)}{\partial t}.\label{momentoconjugado}
\end{equation}
Using the previous results, after some algebra, one can determine the locality structure of the new fermionic field. We start by evaluating the $\mathfrak{b}(t,x)-\boldsymbol{\pi}(t,x^\prime)$ anti-commutator
\begin{eqnarray}
\left\{\mathfrak{b}(t,x),\;\boldsymbol{\pi}(t,x^\prime) \right\} = i\int \frac{d^3p}{(2\pi^3)}e^{i(\p-\p^{\prime})}\frac{1}{4m}\Big(\sum_{i=1,3}\lambda_i(\p)\stackrel{\neg}{\lambda_i}(\p)-\sum_{i=2,4}\lambda_i(-\p)\stackrel{\neg}{\lambda_i}(-\p)\Big),
\end{eqnarray}
so that, bearing in mind relations \eqref{orto1} and \eqref{orto2}, one finds
\begin{eqnarray}
\frac{1}{4m}\Big(\sum_{i=1,3}\lambda_i(\p)\stackrel{\neg}{\lambda_i}(\p)-\sum_{i=2,4}\lambda_i(-\p)\stackrel{\neg}{\lambda_i}(-\p)\Big) = \I,
\end{eqnarray}
yielding
\begin{eqnarray}
\left\{\mathfrak{b}(t,x),\;\boldsymbol{\pi}(t,x^\prime) \right\} = i \delta^3\left(x-x^\prime\right)\I.
\end{eqnarray} 
Consequently, the remaining two equal time anti-commutators vanish
\begin{eqnarray}
\left\{ \mathfrak{b}(t,x),\;\mathfrak{b}(t,x^\prime) \right\}= \left\{ \boldsymbol{\pi}(t,x),\;\boldsymbol{\pi}(t,x^\prime) \right\} = 0, 
\end{eqnarray} 
thus, the field $\mathfrak{b}$ is local.

It is worth mentioning that, given the similarities with the Elko spinors \cite{mdobook}, and also with the recent spinors defined in \cite{dharamnewfermions}, the spinors introduced here have essentially the same Hamiltonian and zero-point energy.

\section{Concluding Remarks and Outlooks}\label{remarks}
In this communication, we have shown the possibility to construct an entirely new class of mass-dimension one fermionic field, in which the eigenvectors of the $\Gamma_{\ell}$ play the role of expansion coefficient after a judicious examination of the dual structure. As we may see, the constructed field is local. As firstly noticed in \cite{dharamnewfermions,dharamboson}, such a mechanism may also serve as a general method to explore other classes of the $\Gamma_{\ell}$ set. Such a possibility provides us with new fermions of mass-dimension one, as well as new fields altogether \cite{dharamboson}.

Interestingly enough, another important consequence concerning the results derived above lies in the fact that the dual of the expansion coefficients defined are not the Dirac dual, and since the propagator essentially depends on the expansion coefficients functions and their dual structures, a crevice opens, and one can evade the Weinberg no go theorem, as firstly reported in \cite{nogo}. By exploring a freedom in the dual structure, as computed in \eqref{newdual1} and \eqref{newdual2}, requiring the Lorentz invariance of the norm and also the Lorentz invariance of the spin-sums, such a mechanism show the results are covariant only under a subgroup of Lorentz \cite{ahluwaliahorvath}, not being in conflict with Weinberg works.

These new types of mass-dimension one fermions, as well as other fields in general, still have an unknown physics, which deserves to be carefully explored in detail from a dynamical perspective, and in various scenarios (such as cosmology and phenomenology). The only restriction for the possible interactions, is driven by the argument of power counting re-normalisability and gauge symmetry.


\appendix

\section{Further Investigations on $\Gamma_{\ell}$}
A parenthetic remark, it is possible to verify that momentum-dependent and dual spinors also correspond to eigenvalues of $\Gamma_{12}$. The eigensvectors of $\Gamma_{12}$, in Eqs \eqref{spinors1} and \eqref{spinors2}, are defined in the rest frame referential. If one wish to obtain the ``momentum dependent'' eigenvectors of $\Gamma$, one need to analyse the following
\begin{eqnarray}
\Gamma_{\ell}\lambda_i(0) = \pm\lambda_i(0),  \label{eqA1}
\end{eqnarray}  
where ``$\pm$'' stands for the $+1$ and $-1$ eigenvalues. Thus, multiplying both sides by Lorentz boost operator, $\kappa$, and inserting the identity $(\kappa^{-1}\kappa)$ between $\Gamma_{12}$ and $\lambda_i(0)$, we have
\begin{eqnarray}
\kappa\Gamma_{12}(\kappa^{-1}\kappa)\lambda_i(0) = \pm\kappa\lambda_i(0), 
\end{eqnarray}  
remembering $\lambda(\textbf{p})=\kappa\lambda_i(0)$, such equation can be written in the following fashion
\begin{eqnarray}
\kappa\Gamma_{12}\kappa^{-1}\lambda(\textbf{p}) = \pm\lambda(\textbf{p}). 
\end{eqnarray} 
Note the eigenvectors of $\kappa\Gamma_{12}\kappa^{-1}$ stands for the momentum dependent spinors. 
\begin{eqnarray}
\kappa\Gamma_{12}\kappa^{-1}\lambda(\textbf{p}) = \pm\lambda(\textbf{p}).
\end{eqnarray}  
Now, for the dual structure relations presented along the text, we have
\begin{eqnarray}
\stackrel{\neg}{\lambda}_{j}(\textbf{p})\gamma_0\mathcal{O}^{\dag}[\kappa\Gamma_{12}\kappa^{-1}]^{\dag}\mathcal{O}^{\dag}\gamma_0 = \pm\stackrel{\neg}{\lambda}_{j}(\textbf{p}).
\end{eqnarray}
The dual stands eigenvector of the $\gamma_0\mathcal{O}^{\dag}[\kappa\Gamma_{12}\kappa^{-1}]^{\dag}\mathcal{O}^{\dag}\gamma_0 $ operator. 

Finally, looking towards develop the same procedure for the dual structure  including $\tau$-deformation, we have for $j=1,3$
\begin{eqnarray}\label{A6}
\tilde{\lambda}_j(\textbf{p})(\I + \mathcal{M}(p, E))\gamma_0\mathcal{O}^{\dag}[\kappa\Gamma_{12}\kappa^{-1}]^{\dag}\mathcal{O}^{\dag}\gamma_0(\I + \mathcal{M}(p, E))^{-1} = \tilde{\lambda}_j(\textbf{p}),
\end{eqnarray}
and for $j=2,4$
\begin{eqnarray}\label{A7}
\tilde{\lambda}_j(\textbf{p})(\I - \mathcal{M}(p, E))\gamma_0\mathcal{O}^{\dag}[\kappa\Gamma_{12}\kappa^{-1}]^{\dag}\mathcal{O}^{\dag}\gamma_0(\I - \mathcal{M}(p, E))^{-1} = -\tilde{\lambda}_j(\textbf{p}),
\end{eqnarray}
Note that the duals stands for eigenvectors of the $(\I + \mathcal{M}(p, E))\gamma_0\mathcal{O}^{\dag}[\kappa\Gamma_{12}\kappa^{-1}]^{\dag}\mathcal{O}^{\dag}\gamma_0(\I + \mathcal{M}(p, E))^{-1}$ and $(\I - \mathcal{M}(p, E))\gamma_0\mathcal{O}^{\dag}[\kappa\Gamma_{12}\kappa^{-1}]^{\dag}\mathcal{O}^{\dag}\gamma_0(\I - \mathcal{M}(p, E))^{-1}$ operators, once spinors $\lambda_1$ and $\lambda_3$ hold a similar (but not the same) dual structure when compared with $\lambda_2$ and $\lambda_4$, thus, the relations \eqref{A6} and \eqref{A7} are a direct consequence of equations \eqref{newdual1} and \eqref{newdual2}.

\bibliographystyle{unsrt}
\bibliography{refs}

\end{document}